\documentstyle[12pt,amsmath,amsfonts, amssymb, amsthm, graphicx, hyperref]{article}
\newcommand{\lC}{\mathrm{l\hspace{-2.1mm}C}}
\newcommand{\lR}{\mathrm{I\hspace{-0.7mm}R}}

\numberwithin{equation}{section}

\begin{document}

\hoffset = -1truecm \voffset = -2truecm

\begin{titlepage}
{\flushright \today
\\}
\vskip 1truecm {\center \LARGE \bf Some Aspects of  Spherical
Symmetric Extremal Dyonic Black Holes in $4d \;\, N=1$
Supergravity \\}

\vskip 2truecm

\begin{center}
{\bf
 Bobby E. Gunara$^{\flat, \ddag}$, Freddy P. Zen$^{\flat, \ddag}$, Fiki T. Akbar$^{\ddag}$,\\
 Agus Suroso$^{\ddag}$, and Arianto$^{\flat, \sharp *}$
 \footnote{email: bobby@fi.itb.ac.id, fpzen@fi.itb.ac.id,
 agussuroso@s.itb.ac.id, ft-akbar@s.itb.ac.id}}
\end{center}

\vskip 1truecm

\begin{center}
%{\bf Draft}
$^{\flat}$\textit{Indonesia Center for Theoretical and
Mathematical Physics (ICTMP),
\\
and \\
$^{\ddag}$ Theoretical  Physics Laboratory, \\
Theoretical High Energy Physics and Instrumentation Research
Group,\\
Faculty of Mathematics and Natural Sciences, \\
Institut Teknologi Bandung \\
Jl. Ganesha 10 Bandung 40132, Indonesia.\\
$^{\sharp}$Department of Physics, Udayana University\\
Jl. Kampus Bukit Jimbaran Kuta-Bali 80361, INDONESIA.\\
$^{*}$ Deceased }

\end{center}

%\vskip 1truecm

{\center \large \bf ABSTRACT
\\}
\vskip 0.5truecm

\noindent In this paper we study several aspects of extremal
spherical symmetric black hole solutions of four dimensional $N=1$
supergravity coupled to vector and chiral multiplets with the
scalar potential turned on. In the asymptotic region the complex
scalars are fixed and regular which can be viewed as the critical
points of the black hole and the scalar potentials with vanishing
scalar charges. It follows that the asymptotic geometries are of a
constant and non-zero scalar curvature which are generally not
Einstein. These spaces could also correspond to the near horizon
geometries which are the product spaces of a two anti-de Sitter
surface and the two sphere if the value of the scalars in both
regions coincides.  In addition, we prove the local existence of non-trivial radius dependent  
complex scalar fields which interpolate between the horizon and the asymptotic region.  
We finally give some simple $\,\, {\lC}^{n}$-models with both linear superpotential and gauge
couplings.\\

%%%%
\end{titlepage}

%\vskip 2truecm

%\maketitle

%%%%%%%-Main Body-%%%%%%%%%%%%%%%%

\section{Introduction}

Solitonic solutions such as black holes of $N\geq 2$ supergravity
have been studied and developed over a decade, see for a review
for example in~\cite{Andrianopoli:2006ub}. The main interest of
the study is due to the so-called \textit{attractor mechanism}
which was firstly discovered in four dimensional ungauged $N=2$
supergravity by some authors \cite{Ferrara:1995prd, Ferrara:1995prd1, Ferrara:1995prd2, Ferrara:1995prd3, Ferrara:1997npb} and in
 $N=1$ supergravity without introducing the
scalar potential \cite{Andrianopoli:2007jhep}. The formalism is
basically to find nondegenerate critical points of the black hole
potential $V_{\mathrm{BH}}$ with $V_{\mathrm{BH}} > 0$ and
particularly, with all eigenvalues of the Hessian matrix of
$V_{\mathrm{BH}}$ at its nondegenerate critical points are
strictly positive.\\
\indent In this paper we present some results of a
particular class of black holes in four dimensional $N=1$
supergravity coupled to vector and chiral multiplets in the
presence of a scalar potential $V$. The black hole is
non-supersymmetric and simply admits a spherical symmetry. Since
the theory is coupled to vector and chiral multiplets, it has
electric, magnetic, and scalar charges \cite{Ferrara:1995prd, Ferrara:1995prd1, Ferrara:1995prd2, Ferrara:1995prd3,
Gibbons:1996prl}. Such a black hole can be regarded as a solution
of a set of equations of motions such as the Einstein field
equation, the gauge field and the scalar field equations of
motions by varying the $N=1$ supergravity action with respect to
the metric, gauge fields, and scalar fields on the spherical
symmetric metric.\\
\indent Our main interest is to consider a special class
of black holes, namely extremal black holes. The word ``extreme"
means that the two black hole horizons coincide on which the black
hole potential extremizes at fixed values of scalars.  Such a case
has been considered for supersymmetric black holes in the context
of four dimensional $N=2$ supergravity \cite{Ferrara:1995prd, Ferrara:1995prd1, Ferrara:1995prd2, Ferrara:1995prd3,Gibbons:1996prl, Kallosh:1996, Behrndt:1996, Behrndt:19961} whose asymptotic
background is flat. While, in our case we study extremal
non-supersymmetric black holes with curved asymptotic backgrounds,
{\textit{i.e.}} four dimensional spacetimes of constant (Ricci)
scalar curvature in the $N=1$ theory with non-zero $V$
\footnote{This non-supersymmetric class of black holes has also
been studied in four dimensional $N=2$ supergravity coupled to
vector multiplets with FI terms \cite{Belluci:2008prd} and
recently in \cite{Kimura:2010jhep}.}.\\
\indent Let us mention the results as follows. First, in the
asymptotic region the scalars are frozen with respect to the radial
coordinate $r$ and can be viewed as the critical points
of both the black hole potential $V_{\mathrm{BH}}$ and the scalar
potential $V$ in order to have a regular value of the scalars. 
The geometries are of a constant scalar curvature which are 
neither Einstein nor symmetric space.\\
\indent Second, near the horizon the scalars are also frozen
with respect to $r$ which are the critical points of the so-called
effective black hole potential $V_{\mathrm{eff}}$ which is a
function of both the black hole potential $V_{\mathrm{BH}}$ and
the scalar potential $V$. Here, the black hole geometries are the
product of a two dimensional surface $M^{1,1}$ and the two-sphere
$S^2$. This $M^{1,1}$ must be $AdS_2$ with different radii
compared to $S^2$. Therefore, the near-horizon geometry is not
conformally flat.\\
\indent Third, in order to have a consistent picture we have to extremize 
the ADM mass of extremal black holes \cite{Gibbons:1996prl} 
\footnote{ This ADM mass has been considered in the asymptotic flat case
\cite{Gibbons:1996prl}.}. This setup implies that the scalar charges vanish. So, we could 
identify that the frozen scalars should be identical in both
regions, namely near the horizon and in the asymptotic region. Thus, if the radius of $AdS_2$ is less
than the radius of $S^2$, then the asymptotic geometry has a
negative scalar curvature. Whereas, we have a positive scalar
curvature space in the asymptotic if the radius of $AdS_2$ is
greater than the radius of $S^2$.\\
\indent To complete our analysis, we prove the local existence of non-trivial radius dependent  
complex scalar fields which interpolate between the horizon and the asymptotic region. 
In the case at hand, we simply show that scalar field equations 
of motions satisfy the local Lipshitz condition. This method has been applied in the case of $N=1$
 supersymmetric Yang-Mills with general couplings  \cite{AGTZ} and references therein.\\
\indent The structure of the paper is as follows. Section
\ref{sec:sugra} is  a review of $N=1$ supergravity
coupled to vector and chiral multiplets. Our convention here
follows rather closely~\cite{D'Auria:2001kv, Andrianopoli:2001zh}.
In Section \ref{sec:EOM} we derive the equation of motions
of each field mentioned above. We discuss some aspects of spherical symmetric black holes in 
section 4 which is splited into four parts: in the first part we consider all equations of motions on static 
spherical symmetric metric, in  the second and the third parts we discuss 
some properties of spherical symmetric extremal black hole in the asymptotic region and near the horizon, while  in the 
fourth part we prove the local existence of non-trivial radius dependent  
complex scalar fields.  We give some simple models, 
namely $\,\, {\lC}^{n}$-models with both linear superpotential and gauge couplings 
in Section \ref{sec:model}. Finally, our conclusions is in Section \ref{sec:conclu}.

%\newpage

\section{$N=1$ Supergravity Coupled with Vector and Chiral Multiplets}
\label{sec:sugra}

In this section we review shortly four dimensional $N=1$
supergravity coupled to arbitrary vector and chiral multiplets.
Here, we assemble the terms which are useful for our analysis in the
paper. Interested reader can further read, for example, \cite{D'Auria:2001kv, Andrianopoli:2001zh}. The theory
consists of a gravitational multiplet, $n_V$ vector and $n_C$
chiral multiplets. Here, we mention the field content of the
multiplets: a gravitational multiplet  $(e_{\mu}^a, \psi_\mu)$, a
vector multiplet $(A_{\mu}, \lambda)$, and a chiral multiplet $(z,
\chi)$ where $e_{\mu}^a$, $A_{\mu}$, and $z$ are a vierbein, a
gauge field, and a complex scalar, respectively, while $\psi_\mu$,
$\lambda$, and $\chi$ are the fermion fields. The bosonic sector
of the Lagrangian can be written down as \cite{D'Auria:2001kv,
Andrianopoli:2001zh} \footnote{Here, we assume that there is no
volume deformation of K\"ahler manifolds. Such a situation has
also been considered for domain wall cases in
\cite{Gunara:2006ig}. On the other hand, some cases with volume
deformation of K\"ahler manifolds have been studied in several
references \cite{Gunara:2007fw, Gunara:2008if, Gunara:2009tu,
Gunara:2009ve, Gunara:2009, Gunara:2012}.}
\begin{eqnarray}
{\mathcal{L}}^{N=1} &=& -\frac{1}{2}R + {\mathcal{R}}_{\Lambda
\Sigma}\,{\mathcal{F}}_{\mu\nu}^{\Lambda}{\mathcal{F}}^{\Sigma|\mu\nu}
+ {\mathcal{I}}_{\Lambda \Sigma}\,
{\mathcal{F}}_{\mu\nu}^{\Lambda}\widetilde{\mathcal{F}}^{\Sigma|\mu\nu}
\nonumber\\
 && + \, g_{i\bar{j}}(z, \bar{z})\,
\partial_{\mu} z^i \,
\partial^{\mu}\bar{z}^{\bar{j}} - V(z,\bar{z})\;, \label{L}
\end{eqnarray}
where $i, j=1, \ldots, n_c$, $\Lambda, \Sigma=1, \ldots, n_v$, and
$\mu,\nu=0, \ldots, 3$. The quantity $R$ is the Ricci scalar of
four dimensional spacetime, whereas
${\mathcal{F}}_{\mu\nu}^{\Lambda}$ is an Abelian field strength of
$A_{\mu}^{\Lambda}$, and
$\widetilde{{\mathcal{F}}}_{\mu\nu}^{\Lambda}$ is a Hodge dual of
${\mathcal{F}}_{\mu\nu}^{\Lambda}$. Also, we have a
Hodge-K\"ahler manifold ${\bf{M}}$ spanned by the complex scalars
$(z, \bar{z})$ with metric $g_{i\bar{j}}(z, \bar{z}) \equiv
\partial_i \partial_{\bar{j}}K(z,\bar{z})$ where $
K(z,\bar{z})$ is a real function called the K\"ahler
potential.\\
\indent The gauge couplings ${\mathcal{N}}_{\Lambda\Sigma}$ are
arbitrary holomorphic functions, while ${\mathcal{R}}_{\Lambda
\Sigma}$ and ${\mathcal{I}}_{\Lambda \Sigma}$ are the real and
imaginary parts of ${\mathcal{N}}_{\Lambda\Sigma}$, respectively.
Similar to the gauge couplings, the function $W(z)$ is also an
arbitrary holomorphic function called holomorphic superpotential.
The scalar potential $V(z,\bar{z})$ is real and given by
\begin{equation}
 V(z,\bar{z}) = e^{K /M^2_P}\left(g^{i\bar{j}} \, \nabla_i W\,
 \bar{\nabla}_{\bar{j}} \bar{W}
 - \frac{3}{M^2_P} W \bar{W} \right)  \;,
\label{V01}
\end{equation}
where $W$ is a holomorphic superpotential, $K \equiv
K(z,\bar{z})$, and $\nabla_i W\equiv
\partial_i W + (K_i /M^2_P) W$.\\
\indent In addition, the Lagrangian (\ref{L}) has a supersymmetric
invariance with respect to  the variation of fields up to
three-fermion terms \cite{D'Auria:2001kv, Andrianopoli:2001zh}:
\begin{eqnarray}
\delta\psi_{{\bullet}\nu} &=& M_P \left(D_{\nu}\epsilon_{\bullet}
+ \frac{\mathrm{i}}{2}e^{K/2M^2_P}\,W \gamma_{\nu}
\epsilon^{\bullet}
+ \frac{\mathrm{i}}{2M_P} Q_{\nu} \epsilon_{\bullet} \right)\;, \nonumber\\
\delta\lambda_{\bullet}^{\Lambda} &=& \frac{1}{2}
({\mathcal{F}}^{\Lambda}_{\mu\nu}
- {\mathrm{i}}\widetilde{\mathcal{F}}^{\Lambda}_{\mu\nu}) \gamma^{\mu\nu}\epsilon_{\bullet} \;, \nonumber\\
\delta\chi^i &=& {\mathrm{i}}\partial_{\nu} z^i \, \gamma^{\nu}
\epsilon^{\bullet} + N^i \epsilon_{\bullet} \;,\label{susytr}\\
\delta e^a_{\nu} &=&  - \frac{{\mathrm{i}}}{M_P} \, (
\bar{\psi}_{{\bullet}\nu} \, \gamma^a \epsilon^{\bullet}
+ \bar{\psi}^{\bullet}_\nu \, \gamma^a \epsilon_{\bullet} )\;,\nonumber\\
\delta A^{\Lambda}_{\mu} &=& \frac{\mathrm{i}}{2}
\bar{\lambda}_{\bullet}^{\Lambda} \gamma_{\mu}\epsilon^{\bullet} +
\frac{\mathrm{i}}{2} \bar{\epsilon}_{\bullet}
\gamma_{\mu} \lambda^{{\bullet}\Lambda}\;, \nonumber\\
\delta z^i &=& \bar{\chi}^i \epsilon_{\bullet} \;,\nonumber
\end{eqnarray}
where $N^i \equiv e^{K /2M^2_P}\,g^{i\bar{j}}
\bar{\nabla}_{\bar{j}}\bar{W}$,
 $g^{i\bar{j}}$ is the inverse of $g_{i\bar{j}}$,
 and the $U(1)$ connection $Q_{\nu} \equiv  - \left(  K_i
 \,\partial_{\nu}z^i -
K_{\bar{i}}\, \partial_{\nu}\bar{z}^{\bar{i}}\right)$.

\section{The Equations of Motions}
\label{sec:EOM}

Let us first discuss the equations of motions of the fields which
can be obtained by varying the action of the Lagrangian
(\ref{L}) with respect to $g_{\mu\nu}$, $A^{\Lambda}_{\mu}$, and
$z^i$. By setting all fermions vanish at the level of the
equation of motions, we have three equations, namely the Einstein
field equation
\begin{eqnarray}
 R_{\mu\nu} - \frac{1}{2} g_{\mu\nu} R &=& g_{i\bar{j}} \,
 (\partial_{\mu}z^i \partial_{\nu}\bar{z}^{\bar{j}}
 + \partial_{\nu}z^i \partial_{\mu}\bar{z}^{\bar{j}}) - g_{i\bar{j}} \, g_{\mu\nu}
\, \partial_{\rho}z^i \partial^{\rho}\bar{z}^{\bar{j}} \nonumber\\
&& + \, 4 \, {\mathcal{R}}_{\Lambda\Sigma}
\,{\mathcal{F}}_{\mu\rho}^{\Lambda}{\mathcal{F}}^{\Sigma}_{\nu\sigma}
g^{\rho\sigma} - g_{\mu\nu} {\mathcal{R}}_{\Lambda\Sigma}
\,{\mathcal{F}}_{\rho\sigma}^{\Lambda}{\mathcal{F}}^{\Sigma|\rho\sigma}
 + g_{\mu\nu} V \;, \label{Einsteineq}
\end{eqnarray}
the gauge field equation of motion
\begin{equation}
 \partial_{\nu} \left(\varepsilon^{\mu\nu\rho\sigma} \sqrt{-g}
 \, {\mathcal{G}}_{\Lambda|\rho\sigma}\right) = 0\;,
\label{gaugeEOM}
\end{equation}
with
\begin{equation}
{\mathcal{G}}_{\Lambda|\rho\sigma}  \equiv {\mathcal{I}}_{\Lambda
\Sigma} {\mathcal{F}}^{\Sigma}_{\rho\sigma} -
{\mathcal{R}}_{\Lambda \Sigma}
\widetilde{\mathcal{F}}^{\Sigma}_{\rho\sigma} \;,
\label{magnetfield}
\end{equation}
are the electric field strengths, and the scalar field equation of
motion
\begin{equation}
 \frac{g_{i\bar{j}}}{\sqrt{-g}} \, \partial_{\mu} \left( \sqrt{-g} \,g^{\mu\nu}
 \partial_{\nu}\bar{z}^{\bar{j}}
 \right) + \bar{\partial}_{\bar{k}} g_{i\bar{j}} \, \partial_{\nu}\bar{z}^{\bar{j}}
 \partial^{\nu}\bar{z}^{\bar{k}} = \partial_i {\mathcal{R}}_{\Lambda
\Sigma}\,{\mathcal{F}}_{\mu\nu}^{\Lambda}{\mathcal{F}}^{\Sigma|\mu\nu}
+ \partial_i{\mathcal{I}}_{\Lambda \Sigma}\,
{\mathcal{F}}_{\mu\nu}^{\Lambda}\widetilde{\mathcal{F}}^{\Sigma|\mu\nu}
- \partial_i V \;, \label{scalarEOM}
\end{equation}
where $g \equiv {\mathrm{det}}(g_{\mu\nu})$. Additionally, there
are the Bianchi identities
\begin{equation}
 \partial_{\nu} \left(\varepsilon^{\mu\nu\rho\sigma} \sqrt{-g}
 \, {\mathcal{F}}^{\Lambda}_{\rho\sigma}\right) = 0\;,
\label{BianchiId}
\end{equation}
from the definition of ${\mathcal{F}}^{\Lambda}_{\rho\sigma}$.\\
\indent Before proceeding to the explicit model, we  briefly point
out the setup  in this paper as follows. From (\ref{Einsteineq})
we obtain the scalar curvature
\begin{equation}
R = 2 g_{i\bar{j}}  \, \partial_{\mu}z^i
\partial^{\mu}\bar{z}^{\bar{j}}-4 \,V \;, \label{scalarcurv}
\end{equation}
whose dynamics are controlled by $\partial_{\mu}z$ and $z$
together with their complex conjugate. It is easy to see that
the scalar curvature (\ref{scalarcurv}) becomes a constant if the
scalar fields are fixed with respect to the spacetime coordinates,
namely $\partial_{\mu}z^i =0$. In this paper we
consider the case where such situations occur in the asymptotic
and the near horizon regions. To achieve such results, first,
assume that in the asymptotic region the scalar fields have to
be frozen, namely $z^i_0$ and $\partial_{\mu}z^i =0$. Then, there exists
a constant $\kappa$ such that the black hole has a constant scalar
curvature where $\kappa$ is related to the potential $V$ evaluated
at its critical point $(z_0, \bar{z}_0)$. Second, near the horizon, similar as before, the scalar fields are
freezed to $z^i_h$ and $\partial_{\mu}z^i =0$ and $\kappa$ is
related to the so-called effective potential evaluated at its
critical point $(z_h, \bar{z}_h)$ (see section \ref{sec:EBR}). In
the rest of the paper we will construct the model in
which the black holes are spherically symmetric and extremal.

\section{Spherical Symmetric Extremal Black Holes}
\label{sec:SSEBH}
\subsection{General Setup}

\indent Let us start to construct a black hole solution of the
equations (\ref{Einsteineq}), (\ref{gaugeEOM}), and
(\ref{scalarEOM}). Our starting point is the foliowing
ansatz metric
\begin{equation}
ds^2 = e^{A(r)}\, dt^2 - e^{B(r)}\, dr^2 - e^{C(r)}\,
 (d\theta^2 + {\mathrm{sin}}^2\theta \, d\phi^2)\;,
\label{metricans}
\end{equation}
which is static and has a spherical symmetry. Among the functions
$A(r)$, $B(r)$, and $C(r)$, only two of them are independent since
one can redefine the radial coordinate $r$ to absorb one of them.\\
\indent On the ansatz (\ref{metricans}), the next step is to solve
the gauge field equation of motions (\ref{gaugeEOM}) together with
the Bianchi identities (\ref{BianchiId}). By simply taking a case
where the field strength components
${\mathcal{F}}^{\Lambda}_{01}(r)$ and
${\mathcal{F}}^{\Lambda}_{23}(\theta)$ are nonzero, we
obtain
\begin{eqnarray} {\mathcal{F}}^{\Lambda}_{01} &=&
\frac{1}{2} \, e^{\frac{1}{2}(A + B)-C}\,
({\mathcal{R}}^{-1})^{\Lambda\Sigma}({\mathcal{I}}_{\Sigma\Gamma}\,
g^{\Gamma} - q_{\Sigma}) \;,\nonumber\\
{\mathcal{F}}^{\Lambda}_{23} &=& - \frac{1}{2} \, g^{\Lambda} \,
{\mathrm{sin}}\theta  \;, \label{solgaugeEOM}
\end{eqnarray}
where $q_{\Lambda}$ and $g^{\Lambda}$ are the electric and
magnetic charges, respectively \cite{Belluci:2008prd}. Using
(\ref{solgaugeEOM}) we have two sets of equations as follows. The
first set of equations  is coming from the Einstein field equation
and the Maxwell equation, namely
\begin{eqnarray}
- e^{-B}\left( C'' + \frac{3}{4} \, {C'}^2
-\frac{1}{2} \, C'B'
  \right) +  e^{-C} = e^{-B} g_{i\bar{j}}\, z^i{'} \bar{z}^{\bar{j}}{'}
   + V + \, e^{-2C} V_{\mathrm{BH}} \;, \nonumber \\
- \frac{1}{2}\, C'\, \left(\frac{1}{2}\, C' + A'\right) + e^{B-C}
= - g_{i\bar{j}}\, z^i{'} \bar{z}^{\bar{j}}{'} + e^{B} \left(V +
e^{-2C} V_{\mathrm{BH}}\right) \;,
\nonumber\\
 - \frac{1}{2}\,e^{-B}\left(A'' + C'' + \frac{1}{2} (A'+C')(A'-B')
 + \frac{1}{2} \, {C'}^2 \right)
 = e^{-B} g_{i\bar{j}}\, z^i{'} \bar{z}^{\bar{j}}{'} + V - e^{-2C} V_{\mathrm{BH}}  \;,
\label{Einsteineq1}
\end{eqnarray}
where $\nu' \equiv d\nu / dr$, while the second equation is the
scalar field equation of motions given by
\begin{equation}
 g_{i\bar{j}} \, \bar{z}^{\bar{j}}{''} + \bar{\partial}_{\bar{k}}
 g_{i\bar{j}} \, \bar{z}^{\bar{j}}{'} \bar{z}^{\bar{k}}{'}
 + \frac{1}{2} \left(A' - B' + 2C' \right)g_{i\bar{j}} \,
 \bar{z}^{\bar{j}}{'} = e^{B} \left( e^{-2C} \partial_i V_{\mathrm{BH}}
 + \partial_i V \right) \;,\label{scalarEOM1}
\end{equation}
where we have assumed that $z^i$  depend only on the radial
coordinate $r$. The potential $V_{\mathrm{BH}}$ has the form
\begin{eqnarray}
V_{\mathrm{BH}} \equiv - \frac{1}{2} \, (g \:\, q
) \; {\mathcal{M}} \; \left(\begin{array}{c}  g \\
 q \end{array} \right)
 \;,
 \label{VBH}
\end{eqnarray}
which is called the black hole potential \cite{Ferrara:1997npb}
where
\begin{eqnarray}
{\mathcal{M}} = \left(\begin{array}{cc} {\mathcal{R}} +
{\mathcal{I}} \,{\mathcal{R}}^{-1} \,{\mathcal{I}} & -
{\mathcal{I}}\, {\mathcal{R}}^{-1} \\
- {\mathcal{R}}^{-1}\, {\mathcal{I}} & {\mathcal{R}}^{-1}
\end{array} \right) \;. \label{McalBH}
\end{eqnarray}
The function $V$ is the scalar potential (\ref{V01}) and in
addition, $V_{\mathrm{BH}}$ contains all charges, namely electric,
magnetic, and scalar charges, with $V_{\mathrm{BH}}\geq 0$.\\
\indent It is worth mentioning that if the scalars $z$ are fixed
for all $r$, then the black hole geometries indeed have a constant
scalar curvature. This case is nothing but the
Reissner-Nordstr\"om-(anti) de Sitter solution with magnetic
charges.\\
% Such a case is referred to as a double-extreme black hole
%\cite{Gibbons:1996prl, Kallosh:1996}????.\\
%
\indent In the next two sections we show that a regular solution
of (\ref{Einsteineq1}) and (\ref{scalarEOM1}) indeed exists in
particular regions, namely near asymptotic and near horizon
regions. As we will see that around these regions the spacetimes
have constant curvatures demanding that the complex scalar fields
$z^i$ have to be fixed which can be viewed as critical points of
potentials defined in the theory.

\subsection{Black Hole Geometries Near Asymptotic Region}
\label{sec:exdyon}

In this section we construct a special solution of
(\ref{Einsteineq1}) around $r \to +\infty$ in which the scalars
$z$ are frozen and can be viewed as critical points of the black
hole and the scalar potentials. Or in other words we restrict
ourselves to a regular solution of (\ref{scalarEOM1}) in this
region. As mentioned in the preceding section, the scalar
curvature (\ref{scalarcurv}) becomes
constant.\\
\indent Our starting points is to take the condition
\begin{eqnarray}
z^i{'}(r) &\to& 0  \;, \nonumber\\
z^i(r) &\to& z^i_0 \;, \label{extremcon}
\end{eqnarray}
around the asymptotic region. We simply then set
\begin{equation}
C(r) = 2 \ln r  \;,
\end{equation}
since the ansatz metric (\ref{metricans}) admits only two
independent functions among $A(r)$, $B(r)$, and $C(r)$. So, from
(\ref{Einsteineq1}) we find that the geometry of black holes has
the form
\begin{equation}
 ds^2 = \Delta\, dt^2 - \Delta^{-1}\, dr^2 -  r^2\,
 (d\theta^2 + {\mathrm{sin}}^2\theta \, d\phi^2)\;,
\label{metricsol}
\end{equation}
where
\begin{equation}
\Delta \equiv 1 - \frac{2 \eta}{r} + \frac{V^0_{\mathrm{BH}}}{r^2}
- \frac{1}{3} V_0\, r^2 \;, \label{Ar}
\end{equation}
and
\begin{eqnarray}
 V^0_{\mathrm{BH}} &\equiv&  V_{\mathrm{BH}}(z_0, \bar{z}_0) \;, \nonumber\\
 V_0 &\equiv&  V(z_0, \bar{z}_0) \;. \label{Vcon}
\end{eqnarray}
The metric (\ref{metricsol}) has a constant scalar curvature but
not Einstein describing a non-supersymmetric solution since the
variations of the fermionic fields in (\ref{susytr}) do not
vanish. The form of (\ref{metricsol}) looks like
Reissner-Nordstr\"om-(anti) de Sitter metric, and since we are
dealing with asymptotic geometries, (\ref{Ar}) must be strictly
positive. Therefore, it does not poses any positive root in the
region.\\
\indent Let us make the above statements more detail in the case
 $V_0 = 0$ and then defining
\begin{equation}
\Delta_0 \equiv r^2 - 2 \eta r + V^0_{\mathrm{BH}}  \;.
\label{Ar0}
\end{equation}
The first case is when if $\eta > 0$, we have $\eta <
\left(V^0_{\mathrm{BH}}\right)^{1/2}$ and $V^0_{\mathrm{BH}}
>0$. This means that (\ref{Ar0}) does not have any root. On the
other hand, if $\eta < 0$, then we have negative roots of
(\ref{Ar0}).\\
\indent Now let us turn to the scalar field equation of motions
(\ref{scalarEOM1}). In this region, taking into account $z^i{'} =
0$, (\ref{scalarEOM1}) limits to
\begin{equation}
\frac{g^0_{i\bar{j}}}{r}  \, \left(r \bar{z}^{\bar{j}}\right){''}
= \Delta^{-1} \left( \frac{1}{r^4}
  \left(\partial_i V_{\mathrm{BH}}\right)_{z_0}
 + \left(\partial_i V\right)_{z_0} \right) \;,\label{scalarEOM2}
\end{equation}
which gives
\begin{eqnarray}
z^i{'}  &=&  - \frac{\Sigma^i}{r^2} + \left( P{'}(r)
   \left(g^{i\bar{j}}\, \bar{\partial}_{\bar{j}} V_{\mathrm{BH}}\right)_{z_0}
 + Q{'}(r) \left(g^{i\bar{j}} \,\bar{\partial}_{\bar{j}} V\right)_{z_0} \right)
 \;,\nonumber\\
z^i &=& z^i_0 + \frac{\Sigma^i}{r} + \left( P(r)
   \left(g^{i\bar{j}}\,  \bar{\partial}_{\bar{j}} V_{\mathrm{BH}}\right)_{z_0}
 + Q(r) \left(g^{i\bar{j}}\, \bar{\partial}_{\bar{j}} V\right)_{z_0}
 \right)\;, \label{solscalarEOM2}
\end{eqnarray}
where $\Sigma^i$ are the scalar charges introduced in
\cite{Gibbons:1996prl} \footnote{The quantities $\Sigma^i$  are called scalar charges  because the second term in (\ref{solscalarEOM2}) looks like the electrostatic Coulomb potential. They can be viewed as the sources for the moduli but they are not conserved \cite{Gibbons:1996prl}.} . The functions $P(r)$ and $Q(r)$ are
\begin{eqnarray}
P(r) &=& \frac{1}{r}\int \left( \int \frac{\Delta^{-1}}{r^3} \, dr \right) dr \;, \nonumber\\
Q(r) &=& \int \left(\int r \, \Delta^{-1}  \, dr \right) dr \;.
\label{PQtil}
\end{eqnarray}
Since the second term in (\ref{solscalarEOM2}) is suppressed in this limit, so  in order to have a consistent picture with the condition
(\ref{extremcon}) it should be then
\begin{eqnarray}
\left( \partial_i V_{\mathrm{BH}}\right)_{z_0} &=& 0 \;, \nonumber\\
\left( \partial_i V\right)_{z_0}  &=& 0 \;, \label{extremcon1}  \ .
\end{eqnarray}
In other words, the moduli fields $z^i_0$ can be thought of as
critical points of both the scalar and black hole potentials,
namely defined by (\ref{V01}) and (\ref{VBH}) respectively,
describing vacua of the theory. Moreover, the first and the second
conditions in (\ref{extremcon1}) may prevent the scalar fields to
be ill defined caused by (\ref{PQtil}) in the asymptotic region.
This can be easily  showed, for example, when the geometries are
of zero scalar curvature, namely $V_0 = 0$, whose functions $P(r)$
and $Q(r)$ have the form
\begin{eqnarray}
P(r) &=& \frac{1}{V^0_{\mathrm{BH}}} \, {{\mathrm{ln}}} r + ... \;,\nonumber\\
Q(r)  &=& \frac{1}{6} \, r^2 + M r +... \;,\nonumber
\end{eqnarray}
respectively, where the dots represent regular terms as $r \to
+\infty$.\\
\indent Let us proceed by discussing the Komar integral for
asymptotically constant scalar curvature spacetimes. Since the
tensor energy-momentum $T_{\mu\nu}$ does not vanish in general,
the form of the Komar integral should be
\begin{eqnarray}
Q = \int_{\partial\Sigma} dS_{\mu\nu} \left( \nabla^{\mu}
\xi^{\nu} + \omega^{\mu\nu}\right) \ , \label{Komarint}
\end{eqnarray}
such that from Stokes's theorem we have
\begin{equation}
\nabla_{\mu} \omega^{\mu\nu} = - \left( T^{\nu}_{~\mu} -{
\frac{1}{2}} ~\delta^{\nu}_{~\mu} ~T \right) \xi^{\mu} \ .
\label{Komarintcon}
\end{equation}
The function $\omega^{\mu\nu}$ is an antisymmetric 2-form, whereas
$\xi^{\mu}$ is a Killing vector. In our case the surface
$\partial\Sigma$ is clearly the 2-sphere. The new Komar integral
(\ref{Komarint}) generalizes the case of asymptotically Einstein
spacetimes \cite{Kastor:2008cqg} which gives the same Smarr
formula as in the asymptotically flat case
\cite{Gibbons:1996prl}.\\
\indent It is worth mentioning that we could have asymptotically
Einstein spacetime for uncharged black holes, namely de Sitter
($dS_4$) and anti-de Sitter ($AdS_4$) for $V_0 \ne 0$ or
Ricci-flat for $V_0 = 0$. Or one could have asymptotically
symmetric spacetimes, if both $\eta$ or $V^0_{\mathrm{BH}}$
vanish. The scalar fields have already a regular value ensured by
(\ref{extremcon1}) in the region.

\subsection{Black Hole Geometries Near the Horizon} \label{sec:EBR}

This section is devoted to show the existence of
Bertotti-Robinson-like geometries when the scalars are frozen near
the horizon and can be viewed as critical points of an effective
scalar potential which is similar to the case of $N=2$
supergravity \cite{Belluci:2008prd}. These black holes are related
to the attractor model discussed, for example, in
\cite{Andrianopoli:2006ub, Ferrara:1995prd, Ferrara:1997npb}.\\
\indent The first step is to freeze the complex scalars, $z^i{'}
=0$ and correspondingly, the near horizon geometry of the metric
(\ref{metricans}) is a product of two surfaces $M^{1,1} \times
S^2$, where $M^{1,1}$ and $S^2$ are respectively two dimensional
surfaces and two-spheres. The setup then implies that the
functions in (\ref{metricans}) are governed by
\begin{eqnarray}
 \frac{1}{2}\,e^{-B}\left(A''  + \frac{1}{2} A'(A'-B')
  \right) &=& \ell \;, \nonumber\\
C &=& {\mathrm{ln}}r_h \;, \label{horgeomcon}
\end{eqnarray}
where $r_h \equiv r_h(g, q)$ is the radius of $S^2$, while the
first equation in (\ref{horgeomcon}) determines the geometry of
$M^{1,1}$ with $\ell
\equiv \ell(g, q)$. \\
\indent Next, in this limit the equations in (\ref{Einsteineq1})
and (\ref{scalarEOM1}) reduce to
\begin{eqnarray}
\frac{1}{r_h^2} = \frac{1}{r_h^4} V^h_{\mathrm{BH}} + V_h \;,
\nonumber\\
\ell = \frac{1}{r_h^4} V^h_{\mathrm{BH}} - V_h \;, \label{fieldEOMhor}\\
 \left(\frac{1}{r_h^4} \frac{\partial V_{\mathrm{BH}}}{\partial
z^i} + \frac{\partial V}{\partial z^i} \right)(p_h) = 0 \nonumber
\;,
\end{eqnarray}
and $p_h \equiv (z_h, \bar{z}_h)$ where we have introduced
\begin{eqnarray}
V^h_{\mathrm{BH}} &\equiv&  V_{\mathrm{BH}} (p_h) \;,\nonumber\\
V_h &\equiv&  V(p_h) \;, \\
\lim_{r \to r_h} z^i &\to& z^i_h \nonumber \;.
\end{eqnarray}
A set of solutions of (\ref{fieldEOMhor}) is given by
\begin{eqnarray}
r_h^2 &=&  V^h_{\mathrm{eff}} \;,
\nonumber\\
\ell^{-1} &=& \frac{V^h_{\mathrm{eff}}}{\sqrt{1 -4
V^h_{\mathrm{BH}} V_h}}
 \;, \label{solfieldEOMhor}\\
\frac{\partial V_{\mathrm{eff}}}{\partial z^i} (p_h)  &=& 0
\nonumber \;,
\end{eqnarray}
where
\begin{equation}
 V_{\mathrm{eff}} \equiv  \frac{1-\sqrt{1-4 V_{\mathrm{BH}}V}}{2
 V}\;\label{effV}
\end{equation}
is called the effective black hole potential
\cite{Belluci:2008prd} and
\begin{equation}
V^h_{\mathrm{eff}} \equiv V_{\mathrm{eff}}(p_h) \;.
\end{equation}
The last equation in (\ref{solfieldEOMhor}) proves that  the
scalars $z_h$ are indeed the critical points of $V_{\mathrm{eff}}$
in the scalar manifold ${\bf{M}}$ near the horizon and $z_h \equiv
z_h(g,q)$. In this case the black hole entropy simply takes the
form \cite{Bekenstein:1973prd, Bekenstein:1973prd1, Bekenstein:1973prd2}
\begin{equation}
S = \frac{A_h}{4} = \pi r_h^2 = \pi V^h_{\mathrm{eff}}\;.
\label{SBH1}
\end{equation}
In addition, the positivity of the entropy (\ref{SBH1}) restricts
$r^2_h > 0$. From all of the above results it follows that one can get if
$B = \pm A$, then $M^{1,1} \simeq AdS_2$. Whereas, the timelike
condition of the Killing vector $\xi = \frac{\partial}{\partial
t}$ excludes the case of $M^{1,1} \simeq dS_2$ \cite{KLR}.\\
\indent Let us relate these results to the results in the
preceding subsection. Employing the Komar integral
(\ref{Komarint}), we get the ADM mass $M_{ex}(z_0(g,q))$ for
extremal black holes, while  we use covariant methods \cite{HS} to obtain \cite{Gibbons:1996prl}
\begin{equation}
\frac{\partial M_{ex}}{\partial z^{i}}(z_0, \bar{z}_0)=  - g_{i\bar{j}}(z_0, \bar{z}_0) \bar{\Sigma}^{\bar{j} } \ . 
\label{MADMderivative}
\end{equation}
Extremizing (\ref{MADMderivative})  implies $\Sigma^i = 0$ for every $i$. Since our black hole is static with vanishing scalar charges, we would  have  \cite{Gibbons:1996prl}
\begin{equation}
z^i_0 = z^i_h\;, \quad {\mathrm{for \; every}}\, i
\end{equation}
which implies that the scalar curvature (\ref{scalarcurv}) becomes
\begin{equation}
R= -4 V_0 = -4\, (\frac{1}{r_h^2}- \ell)\;.
\end{equation}
As observed in \cite{Belluci:2008prd}, in the first case the
spacetime is not conformally flat since $r_a \ne r_h$ where $r_a
\equiv \ell^{-1/2}$ is the radius of $AdS_2$. If the asymptotic
geometry is the spacetime of negative curvature, then $r_a < r_h$.
While for the case of $r_a > r_h$, the asymptotic geometry has a
positive curvature. The case of asymptotically symmetric spaces
such as $AdS_4$ has been discussed, for example, in
\cite{Kimura:2010jhep}.\\
\indent We give in order some remarks. As
mentioned in the previous section, the black hole potential
$V_{\mathrm{BH}} \ge 0$, while the scalar potential $V$ is not
necessarily positive. Therefore, the effective potential
$V_{\mathrm{eff}}$ takes the real value with necessary condition
\begin{equation}
V_{\mathrm{BH}}V < \frac{1}{4} \;, \label{necessrealcon}
\end{equation}
where the regularity of the first order derivative of the
effective black hole potential (\ref{effV}) forbids the equal
sign. Moreover, in order to get a consistent picture the entropy
(\ref{SBH1}) demands that $V_{\mathrm{eff}}$ must be
strictly positive at the ground states. We also have
\begin{eqnarray}
\lim_{V \to 0} V_{\mathrm{eff}} &=& V_{\mathrm{BH}}
\;,\nonumber\\
 \lim_{V_{\mathrm{BH}} \to 0^+}
V_{\mathrm{eff}} &=& 0 \;. \label{limVeff}
\end{eqnarray}
From (\ref{fieldEOMhor}) and (\ref{solfieldEOMhor}) we can
directly see that the second equation in (\ref{limVeff}) is a
singular case with vanishing entropy (\ref{SBH1}). Another
singular model is when $M^{1,1}$ is flat Minkowski surface
$\lR^{1,1}$.\\
%%%%%
\indent  The behaviour of the scalar fields near the horizon takes the condition
\begin{eqnarray}
\lim_{r \to r_h} z^i & \to & z^i_h \ , \nonumber \\
\lim_{r \to r_h} z^i{'} & \to & 0 \ , \label{extremcon2}
\end{eqnarray}
%%%%%
and the equation (\ref{scalarEOM1}) becomes simply
\begin{equation}
  \bar{z}^{\bar{j}}{''}
= \frac{\ell^{-1}}{(r - r_h)^2} \left( g^{i\bar{j}} \frac{\partial V_{\mathrm{eff}}}{\partial z^i} (p_h) \right) \;,\label{scalarEOM3}
\end{equation}
whose solution is given by
\begin{equation}
  \bar{z}^{\bar{j}}
= \bar{z}^{\bar{j}} _h - \ell^{-1} {\mathrm{ln}} | r - r_h | \left( g^{i\bar{j}} \frac{\partial V_{\mathrm{eff}}}{\partial z^i} (p_h) \right) \ .\label{solscalarEOM3}
\end{equation}
So, in order to have a regular solution near $r = r_h$, the last equation in (\ref{solfieldEOMhor}) must be fulfilled. \\
%%%
\indent In this $N=1$ theory the static solution (\ref{metricans}) does break supersymmetry because it does not exist any central charges. So, it is not possible to have an enhancement of supersymmetry in all regions. This is in contrast to the case of black holes in the $N=2$ theory where the enhancement of supersymmetry occurs near the horizon \cite{CFGK}. \\
%%%%%%%%%%
\indent So far, the class of solutions in this paper does not exist in string theory compactified on Calabi-Yau with fluxes since the ground states of the theory in the asymptotic region should be flat Minkowski or anti-de Sitter (AdS), see for example \cite{Kallosh:2005}. However, near the horizon one can apply  AdS$_2$/CFT$_1$ correspondence to find a precise relation between extremal black hole entropy and degeneracy of black hole microstates \cite{Sen:2009}.

%%%%%%%
\subsection{Local Existence} 
\label{LocExist} 

In this section we show  the local existence of non-trivial radius dependent solutions of the scalar equations of motions (\ref{scalarEOM1}). This class of solutions interpolates between the two regions, namely the horizon and asymptotic regions. The logic of this section follows rather closely \cite{AGTZ}. Interested reader can further consult the reference.\\
\indent First of all, we should put some conditions on K\"ahler geometries. The K\"ahler potential  $K \equiv K(z, \bar{z})$ and the Levi-Civita connection $\Gamma$ should satisfies 
%%%
\begin{eqnarray}
K \leq \Phi(|z|) \: , \nonumber\\
\left|\Gamma\right| \leq |\tilde{\Gamma}| \: ,\label{Kpotsymb}
\end{eqnarray}
where $|z| =
\left(\delta_{i\bar{j}} z^{i} \bar{z}^{\bar{j}}\right)^{\frac{1}{2}}$ and $\tilde{\Gamma}$ is the Christoffel symbol of $\tilde{g}$. In other words, the equations in  (\ref{Kpotsymb}) show that the geometry of the $\sigma$-model bounded above by $U(n_{c})$ symmetric K\"ahler geometries with K\"ahler potential  $\Phi(|z|) $.\\
%%%
\indent Defining 
\begin{equation}
F(|z|) =  \frac{1}{4|z|^{2}}\left(\Phi''-\frac{\Phi'}{|z|}\right)
\end{equation}
with  $\Phi' = d \Phi / d |z|$ and $\epsilon$ is a nonnegative constant, and taking the condition
\begin{equation}
\left|\frac{F'}{2|z|}\right|  \leq  \epsilon \: ,\label{ConditionKahler}
\end{equation}
%%%%%%
 then we have the following estimates
\begin{eqnarray}
\left|K\right| & \leq  & \frac{\epsilon}{6} \left| z \right|^{6} +
\frac{C_{1}}{2}\left| z \right|^{4} + C_{2}\left| z \right|^{2} +
C_{3}  \: , \nonumber \\
\left|\Gamma\right| & \leq & 2\epsilon | z |^{3} + C_{1} | z | \: . \label{KpotGammacon}
\end{eqnarray}
%%%
where $C_1, C_2, C_3$ are real constants. Note that several examples of K\"ahler manifolds such as ${\mathbb{C}}^{n_c}$ and ${\mathbb{C}P}^{n_c}$  satisfy (\ref{ConditionKahler}). For ${\mathbb{C}}^{n_c}$,  $F$ vanishes, and hence $\frac{F'}{2| z |}$ is bounded by 0. In case of ${\mathbb{C}P}^{n_c}$, the K\"ahler potential (using standard Fubini-Study metric) is given by
\begin{equation}
\Phi_{{\mathbb{C}P}^{n_c}}(| z |) = \ln(1+| z |^{2}) \: .
\end{equation}
Then we have 
\begin{equation}
\left|\frac{F'}{2| z |}\right| = \frac{2}{\left(1 + | z |^{2}\right)^{3}} \: ,
\end{equation}
which is bounded above by 2.\\
%%%%
\indent The next step is to assume the functions $e^A $ and $e^B$ to be at least $C^2$ function satisfying
\begin{eqnarray}
| e^B |  & \le & C_4  \ , \nonumber\\
| A' -B'  | & \le & C_5  \: , \label{syaratAB}
\end{eqnarray}
where $C_4, C_5$  are positive constants. Such conditions mean that there is no singularity between the region and are  possible because around the asymptotic region we have
\begin{eqnarray}
| e^B |  & \simeq & O(r^{-2}) \ , \nonumber\\
| A' -B'  | & \simeq & O(r^{-1}) \: .
\end{eqnarray}
%%%%%
The final assumption is that all potentials, namely ${\mathcal{V}}  \equiv (V_{\mathrm{BH}}, V)$ to be at least a $C^{2}$ function and a satisfies the local Lipshitz condition
\begin{equation}
\|\partial_{j} {\mathcal{V}} (\tilde{z})-\partial_{j}{\mathcal{V}}(z)\| \leq C(\| \tilde{z} \|,\| z \|)\|\tilde{z} - z \| \:, \label{ScalarPotentialcond}
\end{equation}
where $\| ~ \|$ means  the square integrable norm of a Sobolev space over $[r_ h , +\infty)  \subseteq \mathbb{R}$ and $C(\| \tilde{z} \|,\| z  \|)$ is a bounded function depend on $\| z \|$. The condition above implies that the holomorphic superpotential has to be at least a $C^{3}$ function.\\
%%%%
\indent Now we can finally discuss the final step of the proof. Firstly, we define a function
%%%%
\begin{equation}
J(u) \equiv  -\bar{\partial}_{\bar{k}}
 g_{i\bar{j}} \, \bar{z}^{\bar{j}}{'} \bar{z}^{\bar{k}}{'}
 - \frac{1}{2} \left(A' - B' + 2C' \right)g_{i\bar{j}} \,
 \bar{z}^{\bar{j}}{'}  +  e^{B} \left( e^{-2C} \partial_i V_{\mathrm{BH}}
 + \partial_i V \right)  \  , \label{Jfunc}
\end{equation}
%%%%
where $u \equiv  (z, \bar{z}, z', \bar{z}')$. Secondly, using the conditions (\ref{KpotGammacon}), (\ref{syaratAB}), and (\ref{ScalarPotentialcond}), and employing a tedious computation similar to \cite{AGTZ} we obtain the local Lipshitz condition
for $J(u)$, namely
\begin{equation}
  \| J (\tilde{u}) - J(u) \| \leq C(\| \tilde{u} \|,\| u \|)\|\tilde{u} - u \|  \ .
\end{equation}

\section{Simple $\; {\mathrm{l\hspace{-2.9mm}C}}^{n_c}$-Models}
\label{sec:model}

In this section we consider some simple models on $\; {\lC}^{n_c}$
whose K\"ahler potential has the form
\begin{equation}
 K(z,\bar{z}) =   \vert z \vert^2 \;, \label{KpotC}
\end{equation}
where $\vert z \vert^2 \equiv \delta_{i\bar{j}} \, z^i
\bar{z}^{\bar{j}}$. Particularly, the gauge couplings and the
superpotential have the form
\begin{eqnarray}
{\mathcal{N}}_{\Lambda\Sigma}(z) &=&
(b_0 + b_i z^i) \delta_{\Lambda\Sigma} \;,\nonumber\\
 W(z) &=& a_0 + a_i z^i \;, \label{gcoupw}
\end{eqnarray}
respectively, with $a_0, a_i, b_0, b_i \in \lR$. The black hole
potential and the scalar potential are given by
\begin{eqnarray}
V_{\mathrm{BH}}(x,y) &=& \left(b_0 + b_i x^i + \frac{(b_j
y^j)^2}{(b_0 + b_i x^i)} \right) g^2 - \frac{2b_j y^j}{(b_0 + b_i
x^i)} \, gq + \frac{q^2}{(b_0 + b_i x^i)}   \;,\nonumber\\
 V(x,y) &=& e^{(x^2 + y^2)/M^2_P}
 \Bigg\lbrack a^2 -  \frac{3 a_0^2 }{M^2_P} - \frac{4a_0}{M^2_P} \, a_i x^i
- \frac{1}{M^2_P}  \Big((a_i x^i)^2 + (a_i y^i)^2 \Big) \nonumber\\
&&  + \frac{1}{M^4_P}\left( x^2 + y^2 \right) \Big( (a_0 + a_i
x^i)^2 + (a_i y^i)^2 \Big) \Bigg\rbrack \;, \label{potmod}
\end{eqnarray}
respectively, where we have introduced coordinates $x^i, y^i \in
\lR$ such that $z^i = x^i + {\mathrm{i}} y^i$ and defined some
quantities
\begin{eqnarray}
g^2 \equiv \delta_{\Lambda\Sigma} \, g^{\Lambda}g^{\Sigma} \;, \;
gq
\equiv g^{\Lambda}q_{\Lambda} \;, \nonumber\\
q^2 \equiv \delta^{\Lambda\Sigma} \, q_{\Lambda}q_{\Sigma} \;, \;
a^2 \equiv \delta^{ij}\, a_i a_j \;, \\
x^2 \equiv \delta_{ij}\, x^i x^j \;, \; y^2 \equiv \delta_{ij}\,
y^i y^j \;.\nonumber
\end{eqnarray}
\indent We begin the construction by taking $b_i = 0$ for all $i$
which means for all $z$ we have $\partial_i
{\mathcal{N}}_{\Lambda\Sigma}(z) = 0$. From this it follows that the black
hole potential $V_{\mathrm{BH}}$ becomes
\begin{equation}
V^0_{\mathrm{BH}} = b_0 \, g^2 + \frac{q^2}{b_0} \;,
\label{VBHmodRN}
\end{equation}
which is positive with $b_0 >0$. Firstly, we simply take $z_0 = 0$
and then, get
\begin{equation}
\partial_i V (0) =-\frac{4}{M_P^2} \, a_0 \, a_i = 0 \;,
\label{dV=0}
\end{equation}
which can be split into two cases as follows. The first case is
$a_i = 0$ for all $i$ and $a_0 \ne 0$. The scalar potential
(\ref{V01}) then becomes
\begin{equation}
 V (0) = -\frac{3a^2_0}{M_P^2}  \;, \label{Vads}
\end{equation}
which shows that the scalar curvature of the black hole is
negative. The effective potential (\ref{effV}) in this case
is given by
\begin{equation}
 V_{\mathrm{eff}} (0) = \frac{M_P^2}{6 a^2_0}
 \left(\sqrt{1 + \frac{12 a^2_0 b_0 }{M_P^2}
 \left(g^2 + \frac{q^2}{b^2_0} \right)} \label{effVads}
 - 1\right) \;,
\end{equation}
which is strictly positive. The Hessian matrix of the scalar
potential (\ref{V01}) is simply
\begin{equation}
\partial_i \bar{\partial}_{\bar{j}} V_{\mathrm{eff}}(0) =
    - \frac{4 a_0^2}{M^4_P} \, \delta_{i\bar{j}}  \,
    \frac{\partial V_{\mathrm{eff}} }{\partial V}(0) \;,
\end{equation}
with ${\partial V_{\mathrm{eff}} }/{\partial V}(0) > 0$ showing
that the model is not attractive.\\
\indent The second case is $a_i \ne 0$ for some $i$ and $a_0 = 0$
which follows that the scalar potential (\ref{V01}) is simply
\begin{equation}
 V (0) =  a^2  \;,
\end{equation}
ensuring that the black hole has a positive scalar curvature
background. The Hessian matrix of the effective potential
(\ref{effV}) has the form
\begin{equation}
\partial_i \bar{\partial}_{\bar{j}}V_{\mathrm{eff}} (0) = \frac{2}{M^2_P}
\left( a^2 \, \delta_{ij} - \, a_i \, a_j \right)  \,
\frac{\partial V_{\mathrm{eff}} }{\partial V}(0) \;.
\label{HessV1}
\end{equation}
In this model there may exist an attractor if all eigenvalues of
(\ref{HessV1}) are strictly positive but $n_c \ne 1$.\\
 \indent Finally, we consider a more general case, namely
\begin{eqnarray}
\partial_i V_{\mathrm{BH}} &=& 0 \quad {\mathrm{and}}
\quad \partial_i {\mathcal{N}}_{\Lambda\Sigma} \ne 0
\;,\nonumber\\
\partial_i V  &=& 0 \;. \label{extremcon3}
\end{eqnarray}
Here, we simply set $n_c = 1$, but $n_v > 1$. Moreover, $a_1 = 0$,
while the other pre-coefficients are non-zero. After some steps,
we find that the critical point is
\begin{eqnarray}
 x_0 &=& - \frac{b_0}{b_1} + \frac{1}{b_1 g^2}
 \sqrt{ g^2  q^2 - (gq)^2} \;,\nonumber\\
y_0 &=& \frac{gq}{b_1 g^2}  \;, \label{critpoint}
\end{eqnarray}
with
\begin{eqnarray}
b_1 = \frac{1}{M_P \sqrt{2}}\left( (gq)^2 g^{-4} + \left(-b_0 +
g^{-2} \sqrt{ g^2 q^2 - (gq)^2} \right)^2  \right)^{1/2} \;.
\label{b1}
\end{eqnarray}
In the case at hand, the potentials in (\ref{potmod}) have the
form
\begin{eqnarray}
V_{\mathrm{BH}}(x_0, y_0) &=&  2\, \sqrt{ g^2  q^2 - (gq)^2} \;,\nonumber\\
 V (x_0, y_0) &=& -\frac{e^2 a^2_0 }{M_P^2} \;, \label{potmod1}
\end{eqnarray}
and thus, we have a dyonic black hole with negative scalar
curvature and positive black hole potential since $g^2  q^2 >
(gq)^2$. The effective potential (\ref{effV}) in this model has
the form
\begin{equation}
 V_{\mathrm{eff}} (x_0, y_0) = \frac{M_P^2}{e^2 a^2_0}
 \left(\sqrt{1 + \frac{8 e^2 a^2_0  }{M_P^2}
 \sqrt{ g^2  q^2 - (gq)^2}} \label{effVads1}
 - 1\right) \;.
\end{equation}
The analysis of the Hessian matrix of (\ref{effV}) at $(x_0, y_0)$
shows that this model admits an attractor since ${\partial
V_{\mathrm{eff}} }/{\partial V}(x_0, y_0) > 0$ and ${\partial
V_{\mathrm{eff}} }/{\partial V_{\mathrm{BH}}}(x_0, y_0) > 0$.

\section{Conclusions}
\label{sec:conclu}

In the present paper we have considered several aspects of
extremal dyonic black holes in four dimensional $N=1$ supergravity
that have electric and magnetic charges with curved asymptotic
backgrounds which are not Einstein spaces. The black holes are
particularly non supersymmetric and spherical symmetric.\\
\indent In the asymptotic region we set the scalars to be fixed,
namely $z^i_0$, which can be viewed as the critical points of the
black hole potential $V_{\mathrm{BH}}$ and the scalar potential
$V$. The black hole geometry tends to have a constant and non-zero scalar
curvature.\\
\indent At the horizon the ansatz metric (\ref{metricans}) becomes
a product of two surfaces at vacua defined in the last equation in
(\ref{solfieldEOMhor}), namely $M^{1,1} \times S^2$. These vacua
correspond to the near-horizon limits of (\ref{Einsteineq1}) and
(\ref{scalarEOM1}) where the scalars $z^i$ are frozen and can be
regarded as critical points of $V_{\mathrm{eff}}$ with additional
condition $V^h_{\mathrm{eff}} > 0$ coming from the positivity of
the entropy (\ref{SBH1}). The surface $M^{1,1}$ is $AdS_2$ which
can be easily seen by taking $B = \pm A$. In general, this
spacetime is not conformally flat since $\ell \ne r^{-2}_h$. In
particular, if all the Hessian eigenvalues of $V_{\mathrm{eff}}$
are strictly positive, then the critical point $p_h$ is an
attractor. Note that we exclude the singularities of this model,
namely $V_{\mathrm{BH}} \to 0^+ $ and the two dimensional surface
$M^{1,1}$ is a flat Minkowskian.\\
\indent Furthermore, the extremal condition (\ref{extremcon1}), extremizing the effective scalar potential (\ref{effV}), 
and setting the scalar charges $\Sigma^i$ vanishes for every $i$ follow that we have to identify $z^i_0 = z^i_h$ for every $i$ which
has been observed previously for asymptotic flat cases
\cite{Gibbons:1996prl}. If the asymptotic geometry
has a negative scalar curvature, then $r_a < r_h$ where $r_a
\equiv \ell^{-1/2}$ is the radius of $AdS_2$. While for the case
of $r_a > r_h$, the asymptotic geometry has a positive scalar
curvature. In both cases they are generally not Einstein.\\
\indent We also have shown the local existence of radial dependent scalar fields 
between the black hole horizon and the asymptotic region. Such a case exists if 
the K\"ahler geometry is bounded above by $U(n_c)$-symmteric  K\"ahler geometry 
satisfying (\ref{KpotGammacon}),  all functions in the metric anstaz (\ref{metricans})  
should be $C^2$ functions satisfying (\ref{syaratAB}), and
the first derivative of all potentials, namely ${\mathcal{V}}  \equiv (V_{\mathrm{BH}}, V)$  should fulfill 
the local Lipshitz condition (\ref{ScalarPotentialcond}).\\
%%%
%
\indent At the end, we have worked out $\; {\lC}^{n_c}$ models in
which the superpotential and the gauge couplings both have the
linear forms. In the the first model where the ground state is
simply the origin for $\partial_i {\mathcal{N}}_{\Lambda\Sigma}(z)
= 0$ case, we have a black hole which is asymptotically space of negative scalar curvature 
and not attractive for the case $a_i =0$ for all $i$ and $a_0 \ne 0$, whereas
for $a_i \ne 0$ for some $i$ and $a_0 = 0$, the black hole has a
negative scalar  and may have an attractor if all eigenvalues
of (\ref{HessV1}) are strictly positive but $n_c \ne 1$. Secondly,
for $\partial_i {\mathcal{N}}_{\Lambda\Sigma}(z) \ne 0$ case, and
simply taking $n_c =1$ we obtain a black hole with
negative scalar curvature which is attractive.\\

\vskip 1truecm

\hspace{-0.2 cm}{\Large \bf Acknowledgement}
\\
\vskip 0.15truecm \hspace{-0.6 cm} \noindent One of us, B.E.G,
would like to thank T. Kimura and P. Smyth for valuable
discussions. He is also grateful to J. Louis and the people at II.
Institut f\"ur Theoretische Physik, Universit\"at Hamburg  for
warmest hospitality during his stay. We also greatly thanks M. Satriawan
for proof reading the manuscript and correcting some grammar.  
This paper was initially funded by ITB
Alumni Association Research Grant (HR IA-ITB) 2009 No.
180a/K01.7/PL/2009 and IMHERE DIKTI Project 2010, ITB Alumni
Association Research Grant (HR IA-ITB) 2010 No.
1443b/K01.7/PL/2010, ITB Alumni Association Research Grant (HR
IA-ITB) 2011  No. 2208c/I1.C01/PL/2011, and then, is extended
by Riset Desentralisasi DIKTI-ITB 2012 No. 003.8/TL-J/DIPA/SPK/2012 and Riset KK-ITB 2012 No. 399/I.1.C01/PL/2012.

\end{document}